\documentstyle[11pt,epsfig]{article}
\title{\bf Unification of Gravity and Electromagnetism Revisited}
\author{A. Borzou\thanks{email:ahmadborzou@yahoo.com}$\,$ and H. R. Sepangi
\thanks{email:hr-sepangi@sbu.ac.ir},\,\,\\
{\small Department of Physics, Shahid Beheshti University, Evin,
Tehran 19839, Iran}}
\begin{document}
\maketitle

\begin{abstract}
Within the context of a $5D$ space-time, we construct a unified theory of gravity
and electromagnetism from which the Einstein field equations and Maxwell
equations emerge, with homogenous Maxwell equations appearing naturally. We
also introduce a well-defined five dimensional energy-momentum tensor
consistent with our unification scheme. A correction term appears in Maxwell
equations which can be used to explain the recently discovered galactic
magnetic fields.
\end{abstract}
\vspace{2cm}

\section{Introduction}
After Einstein published the general theory of relativity in 1915,
numerous attempts were made to generalize the theory in such a way
as to encompass both gravitation and electromagnetism in a unique
geometrical structure. Three historically important such
generalizations which are relevant to our present work here are the
Einstein-Schrodinger theory, the Weyl Geometry and the Kaluza-Klein
theory, briefly reviewed below.

The Einstein-schrodinger Theory \cite{1} is a generalization of vacuum general
relativity which allows for non-symmetric fields. Such a theory, without a
cosmological constant, was first proposed by Einstein and Straus \cite{5, 6, 7,
8, 9}. Later on Schrodinger showed that it could be derived from a very simple
Lagrangian density if a cosmological constant was included \cite{10, 11, 12}.
Einstein and Schrodinger suspected that the theory might include
electrodynamics, but no Lorentz force was found \cite{15, 16} when using the
Einstein-Infeld-Hoffmann (EIH) method \cite{15,16b}. For detail study of
non-symmetric fields take a look at \cite{9}.

The Weyl Geometry \cite{2} came into being after Einstein put forth
his general theory of relativity, which provided a geometrical
description of gravitation. Later on, Weyl proposed a more general
setup which also included a geometrical description of
electromagnetism. In the case of general relativity one has a
Riemannian geometry with a metric tensor $g_{\mu\nu}$ \cite{2aa}. If
a vector undergoes a parallel displacement in this geometry, its
direction may change, but not its length. In Weyl geometry, there is
a given vector $k^{\mu}$ which, together with $g_{\mu\nu}$,
characterizes the geometry. For any given vector $\xi^{\mu}$
undergoing parallel displacement in this geometry, not only the
direction but also the length $\xi$ may change and this change
depends on $k_{\mu}$ according to the relation
\begin{equation}
d\xi=\xi k_{\mu} dx^{\mu}, \label{eq1}
\end{equation}
or
\begin{equation}
\xi=\xi_{0}e^{\int{k_{\mu} dx^{\mu}}}, \label{eq2}
\end{equation}
where $\xi_{0}$ is the length of the original vector before
displacement. The change in the length of $\xi$ in going from one
point to another depends on the path followed, i.e., length is not
integrable. Mathematically the Weyl geometry is rich but it seems
not to coincide with  nature. In general relativity and other main
theories of physics, measuring tools are the same for every event.
Therefore, We have to use the same clocks and rulers in our
experiments. In the Weyl geometry one can arbitrarily regauge the
measuring tools for every event. Einstein is among those who have
objected the theory.  A very detail review on the Weyl geometry can
be found in \cite{2a}.

The Kaluza-Klein (KK) theory \cite{3a, 3aa, 3, 4} is perhaps the
most successful of any generalization of general relativity.
Historically, the unification of gravity and electromagnetism was
first addressed in this model by applying Einstein's general theory
of relativity to a five, rather than four-dimensional space-time
manifold. The field equations would logically be expected to be
$\hat{G}_{AB}=k\hat{T}_{AB}$ in $5D$ with some appropriate coupling
constant $k$ and a $5D$ energy-momentum tensor. However, since the
latter is unknown, from the time of Kaluza and Klein onward much
attention has been paid to the vacuum form of  the field equations
$\hat{G}_{AB}=0$, where $\hat{G}_{AB}=\hat{R}_{AB}-\frac{1}{2}
\hat{R} \hat{g}_{AB}$ is the Einstein tensor, $\hat{R}_{AB}$,
$\hat{R}$ are the five-dimensional Ricci tensor and scalar
respectively, and $\hat{g}_{AB}$ is the five-dimensional metric
tensor\footnote{Capital Latin indices $A$, $B$, $\cdots$ assume the
values 0 $\cdots$ 4, Greek indices run through $0\cdots 3$ and the
five-dimensional Ricci tensor and Christoffel symbols are defined in
terms of the metric as in four dimensions.}. Equivalently, the
defining equations are
\begin{equation}
\hat{R}_{AB}=0, \hspace{1 cm} A, B = 0\cdots 4. \label{eq5}
\end{equation}
These 15  relations serve to determine the 15 components of the
metric $\hat{g}_{AB}$, at least in principle. In practice, this is
impossible without some starting assumption on $\hat{g}_{AB}$.
Kaluza was interested in electromagnetism and realized that
$\hat{g}_{AB}$ can be expressed in the form of a 4-potential
$A_{\alpha}$ which appears in Maxwell's theory. He adopted the
cylinder condition, namely the independence of the components of the
metric to the fifth coordinate, but also assumed
$\hat{g}_{44}=\mbox{const.}$ Here we look at the more general case
where $\hat{g}_{44}=-\phi^2 (x^{\alpha}).$ The coordinates or gauges
are chosen so as to write the $5D$ metric tensor in the form
\begin{equation}
g_{AB}=\left(%
\begin{array}{cc}
  g_{\alpha\beta}-\kappa^2 \phi^2 A_{\alpha}A_{\beta} & -\kappa \phi^2 A_{\alpha} \\
  -\kappa \phi^2 A_{\beta} & -\phi^2 \\
\end{array}%
\right), \label{eq6}
\end{equation}
where $\kappa$ is a coupling constant. The field equations then
reduce to
\begin{equation}
G_{\alpha\beta}=\frac{\kappa^2 \phi^2}{2}
T_{\alpha\beta}-\frac{1}{\phi}(\nabla_{\alpha}\nabla_{\beta}\phi-g_{\alpha\beta}\Box\phi),
\label{eq7}
\end{equation}
\begin{equation}
\nabla^{\alpha}F_{\alpha\beta}=-3\frac{\nabla^{\alpha}\phi}{\phi}F_{\alpha\beta},
\label{eq8}
\end{equation}
and
\begin{equation}
\Box\phi=-\frac{\kappa^2 \phi^3}{4}F_{\alpha\beta}F^{\alpha\beta}.
\end{equation}
Here, $G_{\alpha\beta}$ and $F_{\alpha\beta}$ are the usual $4D$
Einstein and Faraday tensors and $T_{\alpha\beta}$ is the
energy-momentum tensor for an electromagnetic field given by
\begin{equation}
T_{\alpha\beta}=\frac{1}{2}\left(\frac{1}{4}g_{\alpha\beta}F_{\gamma\delta}F^{\gamma\delta}-
F_{\alpha}^{\gamma}F_{\beta\gamma}\right). \label{eq9}
\end{equation}
Equation (\ref{eq7}) gives back the 10 Einstein equations of $4D$
general relativity, but with a right-hand side which in some sense
represents the energy and momentum tensor. Kaluza's case
$\hat{g}_{44}=-\phi^{2}=-1$ together with the identification
$\kappa=(\frac{16\pi G}{c^{4}})^{\frac{1}{2}}$ makes (\ref{eq7})
and (\ref{eq8}) to read
\begin{equation}
G_{\alpha\beta}=\frac{8\pi G}{c^{4}} T_{\alpha\beta},
\end{equation}
and
\begin{equation}
\nabla^{\alpha}F_{\alpha\beta}=0.
\end{equation}
Note that this theory does not contain the two homogeneous Maxwell
equations, $\nabla_{(k}F_{mn)}=0$. Of course, it goes without saying
that the homogeneous Maxwell equations are constraints on the
Faraday tensor and are implicit in the KK theory. However, in what
follows, we show that within the context of the theory presented
here, these equations can be independently obtained form the torsion
free condition, in contrast to the conventional KK theory.

All the studies carried out in this regard in the past  have found a
common problem, namely that the Homogeneous Maxwell equations don
not appear naturally. Also in most five dimensional theories there
is no obvious reason for the cylinder condition. We try to obtain
not only the Einstein and non-homogeneous Maxwell equations but also
homogeneous Maxwell equations and a solution to the latter by only
one assumption. In this paper we investigate a five dimensional
space-time and examine its consequences. This is done by forcing the
connection to have certain properties, thus enabling one to define
the resulting space-time quantities. We show how our assumption
leads to cylinder condition and also how the Einstein field
equations in five dimensions separate into the 4-dimensional
Einstein and Maxwell equations and that the homogeneous Maxwell
equations are included in the theory as $\hat{R}_{(nmk)4}=0$. The
corresponding energy-momentum tensor is then defined and presented.
The conclusions are drawn in the last section.
\section{Five Dimensional Riemannian Space-Time}
Let us start with the line-element for a five dimensional Riemannian space-time
\begin{equation}
d\hat{s}^2=\hat{g}_{AB}d\hat{x}^Ad\hat{x}^B,
\end{equation}
where a hat represents five dimensional objects while objects without it belong to
the ordinary $4D$ space-time. Here $\hat{x}^{\alpha}=x^{\alpha}$ are ordinary
$4D$ coordinates. Capital latin indices also run over 0, 1, 2, 3, and 4.  The
contravariant components of the metric are defined as inverse of the covariant
components
\begin{equation}
\hat{g}_{AB}\hat{g}^{BC}=\delta_A^C.
\end{equation}
In such a space-time, the covariant derivatives are defined as
\begin{equation}
\hat{\nabla}_{A}\hat{f}^{B}=\partial_{A}\hat{f}^{B}+\hat{\Gamma}^{B}_{AC}\hat{f}^{C}.
\end{equation}
Now we can define a geodesic path as the straightest possible curve, i.e. a curve
whose tangent vectors are connected by parallel transport. The tangent vector to
a curve is given by $\hat{\dot{x}}^{A}\hat{e}_{A}$ where $\hat{e}_A$ are a set of
the basis vectors expanding the space and
$\hat{\dot{x}}^{A}=\frac{d\hat{x}^A}{d\tau}$. The equation representing parallel
transportation may then be written as
\begin{eqnarray}
\hat{\dot{x}}^{A}
\hat{\nabla}_{\hat{e}_{A}}\hat{\dot{x}}^{B}\hat{e}_{B}=0,
\end{eqnarray} or
\begin{eqnarray}
{\hat{\ddot{x}}^{B}+\hat{\Gamma}^{B}_{AC}\hat{\dot{x}}^{A}
\hat{\dot{x}}^{C}}=0.\label{ad}
\end{eqnarray}
The $5D$ curvature tensor, $\hat{R}^D_{_{ABC}}$, can be defined as follows
\begin{equation}
\hat{\nabla}_{A}\hat{\nabla}_{B}\hat{f}_{C}-\hat{\nabla}_{B}\hat{\nabla}_{A}\hat{f}_{C}=\hat{R}^D_{_{ABC}}\hat{f}_{D},\label{ab}
\end{equation}
where $\hat{f}_{C}$ is an arbitrary $5D$ vector. Therefore, the
curvature tensor will be as follows
\begin{equation}
\hat{R}^{D}_{_{CBA}}=\partial_{B}\hat{\Gamma}^{D}_{AC}-\partial_{A}\hat{\Gamma}^{D}_{BC}+\hat{\Gamma}^{D}_{B
E}\hat{\Gamma}^{E}_{AC}-\hat{\Gamma}^{D}_{A E}\hat{\Gamma}^{E}_{B
C}\label{a4}.
\end{equation}
The Ricci tensor in a five dimensional Riemannian space-time can
also be defined as follows
\begin{equation}
\hat{R}_{_{CA}}=\hat{R}^{B}_{_{CBA}}\label{a5}.
\end{equation}
Although the 5D Ricci scalar is defined as $\hat{R}=\hat{g}^{CA}
\hat{R}_{_{CA}}$, we do not need it in our 5D field equations any
more. One may now easily write $5D$ field equations as
\begin{equation}
\hat{R}_{_{AB}}-\frac{1}{2} \hat{\Im} \hat{g}_{_{AB}}=\kappa
\hat{T}_{_{AB}},\label{ac}
\end{equation}
where $\hat{R}_{_{AB}}$, $\hat{g}_{_{AB}}$ and $\hat{T}_{_{AB}}$ are
the 5D Ricci, metric, and energy-momentum tensor respectively.
$\hat{\Im}$ is a $5D$ scalar which can be determined using
conservation laws and will be discussed later.
\subsection{The Setup}
At this point, it is appropriate to make our main assumption
\begin{equation}
\hat{\Gamma}^{4}_{AC}=0,\label{a3}
\end{equation}
in the whole $5D$ space-time. Substituting this equation into
(\ref{ad}) leads to the fact that $\hat{\ddot{x}}^4=0$ for particles
that move on geodesics, meaning that for these particles we have
$\hat{\dot{x}}^4=constant$. The implication is that the fifth
component of the position of a particle in our space is the same as
any other. We hope our space-time to include both gravitational and
electromagnetic forces which means that even charged particles in
presence of electromagnetic waves move on geodesics. As a result, in
macroscopic scales one can say that particles never see the fifth
dimension. It is something like the cylinder condition which means
that the derivative with respect to the fifth coordinate on our $4D$
hypersurface is zero. We are now in a position to investigate the
functional form of the Christoffel symbols which we assume are
symmetric in lower indices. Let us begin by considering the well
known fundamental equation
\begin{equation}
\hat{\nabla}_A\hat{g}_{BC}=0,
\end{equation}
which is a result of
\begin{equation}
\partial_A\hat{g}_{BC}=\hat{\Gamma}^{D}_{AB}\hat{g}_{DC}+
\hat{\Gamma}^{D}_{AC}\hat{g}_{BD}.
\end{equation}
We may then write
\begin{equation}
\partial_A\hat{g}_{BC}+\partial_B\hat{g}_{AC}-\partial_C\hat{g}_{AB}=2\hat{\Gamma}^{D}_{AB}\hat{g}_{DC}.
\end{equation}
Therefore, knowing that $\hat{\Gamma}^{4}_{AC}=0$, we find
\begin{equation}
\partial_A\hat{g}_{BC}+\partial_B\hat{g}_{AC}-\partial_C\hat{g}_{AB}=2\hat{\Gamma}^{\lambda}_{AB}\hat{g}_{\lambda C}.\label{aa}
\end{equation}
Before going any further we should make it clear what we mean by the $4D$
metric. If we write the line-element as
\begin{equation}
d\hat{s}^2=\hat{g}_{\alpha
\beta}d\hat{x}^{\alpha}d\hat{x}^{\beta}+2\hat{g}_{\alpha
4}d\hat{x}^{\alpha}d\hat{x}^4+\hat{g}_{4 4}d\hat{x}^4d\hat{x}^4,
\end{equation}
it would be obvious that $\hat{g}_{\alpha \beta}=g_{\alpha \beta}$
are the covariant components of the metric tensor for the four
dimensional space-time, and the contravariant components of the $4D$
metric tensor are defined as
\begin{equation}
g_{\alpha \beta}g^{\beta \gamma}=\delta_{\alpha}^{\gamma}.
\end{equation}
Now going back to our approach to investigate the functional form of the
Christoffel symbols, we replace $C$ in equation (\ref{aa}) with $\sigma$ and find
\begin{equation}
\partial_A\hat{g}_{B \sigma}+\partial_B\hat{g}_{A \sigma}-\partial_{\sigma} \hat{g}_{AB}=
2\hat{\Gamma}^{\lambda}_{AB}\hat{g}_{\lambda
\sigma},
\end{equation}
which leads to what we were looking for
\begin{equation}
\hat{\Gamma}^{\lambda}_{AB}=\frac{1}{2}g^{\lambda
\sigma}(\partial_A\hat{g}_{B \sigma}+\partial_B\hat{g}_{A
\sigma}-\partial_{\sigma} \hat{g}_{AB}). \label{a2}
\end{equation}
One can simply check that when all capital Latin indices are replaced with Greek
indices, then $\hat{\Gamma}^{\lambda}_{\alpha
\beta}=\Gamma^{\lambda}_{\alpha \beta}$, where $\Gamma^{\lambda}_{\alpha
\beta}$ is the ordinary $4D$ connection. This is our final result.  Alternatively, if
$C$ in equation (\ref{aa}) is replaced with $4$, we find
\begin{equation}
\partial_A\hat{g}_{B 4}+\partial_B\hat{g}_{A 4}-\partial_4\hat{g}_{AB}=
2\hat{\Gamma}^{\lambda}_{AB}\hat{g}_{\lambda
4},\label{ae}
\end{equation}
or
\begin{equation}
\partial_A\hat{g}_{B 4}+\partial_B\hat{g}_{A 4}=2\hat{\Gamma}^{\lambda}_{AB}\hat{g}_{\lambda
4}\mid_{on~ the~ hypersurface}.\label{ae}
\end{equation}
Now, taking all the possible values of $A$ and $B$ will arrive at
the following useful results
\begin{equation}
\partial_\alpha \hat{g}_{\beta 4}+\partial_\beta\hat{g}_{\alpha 4}=2\Gamma^{\lambda}_{\alpha \beta}\hat{g}_{\lambda
4}\mid_{on~the~hypersurface},
\end{equation}
\begin{equation}
\partial_\alpha \hat{g}_{4 4}=2\hat{\Gamma}^{\lambda}_{\alpha 4}\hat{g}_{\lambda
4}\mid_{on~the~hypersurface},
\end{equation}
and
\begin{equation}
\hat{\Gamma}^{\lambda}_{4 4}\hat{g}_{\lambda 4}=0\mid_{on~the~hypersurface}.
\end{equation}
Using equations (\ref{a3}) and (\ref{a2}), the Riemann tensor,
(\ref{a4}), will separate into
\begin{equation}
\hat{R}^{\kappa}_{_{CBA}}=\partial_{B}\hat{\Gamma}^{\kappa}_{AC}-\partial_{A}\hat{\Gamma}^{\kappa}_{BC}+\hat{\Gamma}^{\kappa}_{B
\gamma}\hat{\Gamma}^{\gamma}_{AC}-\hat{\Gamma}^{\kappa}_{A
\gamma}\hat{\Gamma}^{\gamma}_{B C}\label{a6},
\end{equation}
and
\begin{equation}
\hat{R}^{4}_{_{CBA}}=0\label{a7}.
\end{equation}
Here again it is clear that $\hat{R}^{\kappa}_{_{\alpha \beta
\gamma}}=R^{\kappa}_{\alpha \beta \gamma}$ where $R^{\kappa}_{\alpha \beta
\gamma}$ is an ordinary $4D$ Riemann tensor. It is now interesting to study the
symmetry properties of the curvature tensor. For this purpose, let us adopt the
geodesic coordinate system in which the curvature tensor can be written as
\begin{equation}
\hat{R}^{D}_{_{CBA}}=\partial_{B}\hat{\Gamma}^{D}_{AC}-\partial_{A}\hat{\Gamma}^{D}_{BC},
\end{equation}
from which we immediately see that
\begin{equation}
\hat{R}^{D}_{_{CBA}}=-\hat{R}^{D}_{_{CAB}}.
\end{equation}
Now, to see if other symmetry properties are satisfied, we  write our equation as
\begin{equation}
\hat{R}_{_{ABCD}}=\hat{g}_{_{AE}}\hat{R}^{E}_{_{BCD}}=
\partial_{_{C}}(\hat{g}_{_{AE}}\hat{\Gamma}^E_{_{BD}})-
\partial_{_{D}}(\hat{g}_{_{AE}}\hat{\Gamma}^E_{_{BC}}),
\end{equation}
showing that unlike the ordinary $4D$ space-time, because of the
unusual characteristics of the connections, the other common
symmetries are not satisfied. In general we have
\begin{equation}
\hat{R}_{_{DCBA}} \neq -\hat{R}_{_{CDBA}} \neq \hat{R}_{_{BADC}}.
\end{equation}
In addition, the first and second Bianchi identities can be written as
\begin{equation}
\hat{R}^{A}_{_{BCD}}+\hat{R}^{A}_{_{DBC}}+\hat{R}^{A}_{_{CDB}}=0,\label{af}
\end{equation}
and
\begin{equation}
\hat{\nabla}_{_{E}}\hat{R}^D_{_{ABC}}+
\hat{\nabla}_{_{C}}\hat{R}^D_{_{AEB}}+\hat{\nabla}_{_{B}}\hat{R}^D_{_{ACE}}=0,
\end{equation}
where both can be verified easily in geodesic coordinate. Since not
all the symmetry properties of  ordinary $4D$ space-time appear in
our theory, naturally not all the known Bianchi identity forms in
$4D$ are correct here. However, there is another form of the First
Bianchi identity which is of much importance to us, namely
\begin{equation}
\hat{R}_{\alpha \beta \gamma 4}+\hat{R}_{\gamma \alpha \beta
4}+\hat{R}_{\beta \gamma \alpha 4}=0\mid_{on~the~hypersurface},\label{ag}
\end{equation}
which can be easily proved in geodesic coordinates using
$\partial_4=0$. The Ricci tensor in this space-time can be obtained
using equations (\ref{a5}), (\ref{a6}), and (\ref{a7}) as
\begin{equation}
\hat{R}_{_{CA}}=\partial_{\kappa}\hat{\Gamma}^{\kappa}_{AC}-\partial_{A}\hat{\Gamma}^{\kappa}_{\kappa
C}+\hat{\Gamma}^{\kappa}_{\kappa
\gamma}\hat{\Gamma}^{\gamma}_{AC}-\hat{\Gamma}^{\kappa}_{A
\gamma}\hat{\Gamma}^{\gamma}_{\kappa C}.
\end{equation}
Again
\begin{equation}
\hat{R}_{\alpha \gamma}=R_{\alpha \gamma},
\end{equation}
where $R_{\alpha \gamma}$ is the ordinary $4D$ Ricci tensor. It is
very simple to show that the Ricci tensor is symmetric.

Now we can investigate the field equations, (\ref{ac}), and
determine the scalar appearing in it. We know that, as mentioned
before, all the matter is confined to our $4D$ hypersurface and
therefore energy-momentum is a conservative quantity in our $4D$
volume. We can interpret $\hat{J}_{A}=\hat{T}_{A
B}\hat{\dot{x}}^{B}$ as current density. Hence the conservation law
can be written as
\begin{equation}
\oint_{\partial\Omega}\hat{J}_{\alpha}df^{\alpha}=
\oint_{\partial\Omega}\hat{T}_{\alpha B}\hat{\dot{x}}^{B}df^{\alpha}=0,
\end{equation}
where $df^{\alpha}$ is the vector representing the area element
perpendicular to the surface $\hat{x}^{\alpha}=const$. Using Gauss's
law, we will have the following equation
\begin{equation}
\int_{\Omega}\nabla^{\alpha}\hat{J}_{\alpha}\sqrt{-g}d^4x=0,
\end{equation}
where $\nabla^{\alpha}$ is the $4D$ covariant derivative. This
integral implies
\begin{equation}
\nabla^{\alpha}\hat{J}_{\alpha}=\nabla^{\alpha}(\hat{T}_{\alpha B}\hat{\dot{x}}^{B})=0,
\end{equation}
or
\begin{equation}
\nabla^{\alpha}(\hat{T}_{\alpha \beta}\hat{\dot{x}}^{\beta}-\hat{T}_{\alpha 4})=0,
\end{equation}
where we use $\hat{\dot{x}}^4=-1$ on the hypersurface. For any
arbitrary but small region we can find vector fields with
$\nabla^{\alpha}\hat{\dot{x}}^{\beta}\simeq0$ \cite{b2}. As a result
\begin{equation}
\nabla^{\alpha}(\hat{T}_{\alpha \beta})\hat{\dot{x}}^{\beta}-\nabla^{\alpha}\hat{T}_{\alpha 4}=0.
\end{equation}
Since $\hat{\dot{x}}^{\beta}$ can be chosen arbitrarily, this equation gives us two
distinct equations
\begin{equation}
\nabla^{\alpha}(\hat{T}_{\alpha \beta})=0,
\end{equation}
and
\begin{equation}
\nabla^{\alpha}\hat{T}_{\alpha 4}=0.
\end{equation}
Now, it is appropriate to take $\hat{T}_{\alpha \beta}$ as the four
dimensional energy-momentum tensor, $T_{\alpha \beta}$, and
$\hat{T}_{\alpha 4}$ as the four dimensional electromagnetic current
vector, $j_{\alpha}$, and write the five dimensional energy-momentum
tensor in the following form
\begin{equation}
\hat{T}_{AB}=\left(%
\begin{array}{ccccc}
  T_{\alpha\beta} &   &   &   & \frac{q}{m}\epsilon j_{0} \\
    &   &   &   & \frac{q}{m}\epsilon j_{1} \\
    &   &   &   & \frac{q}{m}\epsilon j_{2} \\
    &   &   &   & \frac{q}{m}\epsilon j_{3} \\
  \frac{q}{m}\epsilon j_{0} & \frac{q}{m}\epsilon j_{1} & \frac{q}{m}\epsilon j_{2} & \frac{q}{m}\epsilon j_{3} & \hat{T_{44}} \\
\end{array}%
\right),
\end{equation}
where $\epsilon$ is a coupling constant. If we replace $A$ by
$\alpha$ and $B$ by $\beta$, equation (\ref{ac}) will reduce to
\begin{equation}
\hat{R}_{_{\alpha \beta}}-\frac{1}{2} \hat{\Im} \hat{g}_{_{\alpha \beta}}=\kappa
\hat{T}_{_{\alpha \beta}},
\end{equation}
or equivalently
\begin{equation}
R_{_{\alpha \beta}}-\frac{1}{2} \hat{\Im} g_{_{\alpha \beta}}=\kappa
T_{_{\alpha \beta}}.
\end{equation}
Since $\nabla^{\alpha}T_{\alpha \beta}=0$, we have to have
$\nabla^{\alpha}(R_{_{\alpha \beta}}-\frac{1}{2} \hat{\Im} g_{_{\alpha \beta}})=0$
which leads to the following equation
\begin{equation}
 \hat{\Im}=R.
\end{equation}
Here $R$ is $4D$ Ricci scalar. It is easy to show that the $4D$
Ricci scalar is also a scalar in five dimensions. Therefore, we can
write the field equations (\ref{ac}) as follows
\begin{equation}
\hat{R}_{_{AB}}-\frac{1}{2} R \hat{g}_{_{AB}}=\kappa
\hat{T}_{_{AB}}.\label{ac1}
\end{equation}
\section{Electromagnetism}
As the first step  let us derive the Lorentz force which is a part
of the geodesic equation. Knowing that on the geodesic equation
$\hat{\dot{x}}^4=-1$, we can write equation (\ref{ad}) as
\begin{equation}
\hat{\ddot{x}}^{\kappa}+\hat{\Gamma}^{\kappa}_{\alpha \beta}\hat{\dot{x}}^{\alpha}\hat{\dot{x}}^{\beta}-2\hat{\Gamma}^{\kappa}_{\alpha 4}\hat{\dot{x}}^{\alpha}+\hat{\Gamma}^{\kappa }_{4 4}=0.
\end{equation}
The geodesic equation is then given by
\begin{equation}
\hat{\ddot{x}}^{\kappa}+\frac{1}{2}g^{\lambda
\kappa}(\hat{g}_{\lambda\beta,\alpha}+\hat{g}_{\lambda\alpha,\beta}-
\hat{g}_{\alpha\beta,\lambda})\hat{\dot{x}}^{\alpha}\hat{\dot{x}}^{\beta}-g^{\lambda
\kappa}(\hat{g}_{4 \lambda,\alpha}-\hat{g}_{4
\alpha,\lambda})\hat{\dot{x}}^{\alpha
}-\frac{1}{2}g^{\lambda \kappa}\hat{g}_{44,\lambda}=0,\label{eq19}
\end{equation}
noting that partial derivatives with respect to $\hat{x}^{4}$ are
zero, and a comma  indicates partial derivative. We may now define
the vector potential, $A_{\mu}$, as
\begin{equation}
A_{\mu}=\frac{m}{q}\hat{g}_{4 \mu},
\end{equation}
where $m$ and $q$ are the mass and electric charge of our test
particle respectively. Therefore, we define the Faraday tensor as
\begin{equation}
F_{\mu \nu}=\hat{g}_{4 \mu,\nu}-\hat{g}_{4 \nu,\mu},
\end{equation}
where by means of this equation, one can simply find the Lorentz
force in equation (\ref{eq19}).

As was mentioned above and is well known, the homogeneous Maxwell
equations appear as constraints on the Faraday tensor in the KK
theory and are implicitly assumed to hold. They do do not appear in
an independent manner. However, the situation is different in the
theory presented here. It is therefore appropriate at this point to
show that this is indeed the case. To begin with we note that in
classical electrodynamics, the Jacobi identities lead to the
homogenous Maxwell equations if we define our connections as
$D_{\mu}=\partial_{\mu}+A_{\mu}$, where $A_{\mu}$ is the four vector
potential, so that the Faraday tensor is written as $F_{\mu
\nu}=[D_{\mu},D_{\nu}]$. Now, use of the second Jacobi identity
leads to the Bianchi identity in the form
\begin{equation}
[D_{\mu},F_{\nu \lambda}]+[D_{\nu},F_{\lambda
\mu}]+[D_{\lambda},F_{\mu \nu}]=0,
\end{equation}
which is equivalent to
\begin{equation}
D_{\mu}F_{\nu \lambda}+D_{\nu}F_{\lambda \mu}+D_{\lambda}F_{\mu
\nu}=0.
\end{equation}
This is, of course,  the homogenous Maxwell equations \cite{20}. In General
Relativity on the other hand, we replace $D_{\mu}$ by $\nabla_{\mu}$, so that
the first Jacobi identity results in
\begin{equation}
[\nabla_{\mu},R(\partial_{\nu},
\partial_{\lambda})]+[\nabla_{\nu},R(\partial_{\lambda},
\partial_{\mu})]+[\nabla_{\lambda},R(\partial_{\mu},
\partial_{\nu})]=0,
\end{equation}
or
\begin{equation}
R_{\alpha \beta [\mu \nu ;\lambda]}=0.\label{eq100}
\end{equation}
The torsion free condition also provides the first Bianchi identity
\begin{equation}
R_{\lambda [\beta \gamma \delta]}=0.\label{eq101}
\end{equation}
It is now clear that because of the definition of connections in
general relativity, neither the first Bianchi identity, equation
(\ref{eq101}), nor the second Bianchi identity, equation
(\ref{eq100}), lead to homogenous Maxwell equations. Let us now show
that in the model presented here, the first Bianchi identity leads
to the homogenous Maxwell equations. To write the first Bianchi
identity in our model we have to show that
\begin{equation}
\hat{R}^{\alpha}_{\beta \kappa
4}=\frac{1}{2}\nabla_{\kappa}F^{\alpha}_{\beta},\label{eq33}
\end{equation}
where $F_{\alpha \beta}$ is the Faraday tensor in four dimensional
space which is defined by $g_{\mu \nu}$. We write
\begin{equation}
\frac{1}{2}F^{\kappa}_{\sigma}=\frac{1}{2}g^{\kappa
\lambda}(\hat{g}_{4 \lambda,\sigma}-\hat{g}_{4
\sigma,\lambda})=\hat{\Gamma}^{\kappa}_{4\sigma},
\end{equation}
and
\begin{equation}
\frac{1}{2}\nabla_{\kappa}F^{\alpha}_{\beta}=\partial_{\kappa}\hat{\Gamma}^{\alpha}_{4\beta}+
\Gamma^{\alpha}_{\kappa\lambda}\hat{\Gamma}^{\lambda}_{4\beta}-\Gamma^{\lambda}_{\kappa
\beta}\hat{\Gamma}^{\alpha}_{4\lambda}.
\end{equation}
On the other hand we have
\begin{equation}
\hat{R}^{\alpha}_{\beta \kappa
4}=\partial_{\kappa}\hat{\Gamma}^{\alpha}_{4\beta}+\hat{\Gamma}^{\alpha}_{\kappa
\lambda}\hat{\Gamma}^{\lambda}_{4\beta}-\hat{\Gamma}^{\lambda}_{\kappa
\beta}\hat{\Gamma}^{\alpha}_{4\lambda},
\end{equation}
\begin{equation}
\hat{R}^{\alpha}_{\beta \kappa
4}=\partial_{\kappa}\hat{\Gamma}^{\alpha}_{4\beta}+
\Gamma^{\alpha}_{\kappa\lambda}\hat{\Gamma}^{\lambda}_{4\beta}-\Gamma^{\lambda}_{\kappa
\beta}\hat{\Gamma}^{\alpha}_{4\lambda}.
\end{equation}
Hence
\begin{equation}
\hat{R}^{\alpha}_{\beta \kappa
4}=\frac{1}{2}\nabla_{\kappa}F^{\alpha}_{\beta}\label{a8}.
\end{equation}
Now using equation (\ref{a8}), the first Bianchi identity, equation
(\ref{ag}), leads to
\begin{eqnarray}
\frac{1}{2}\nabla_{\kappa}F_{\alpha
\beta}+\frac{1}{2}\nabla_{\beta}F_{\kappa
\alpha}+\frac{1}{2}\nabla_{\alpha}F_{\beta \kappa}=0, \label{eq39}
\end{eqnarray}
showing that, as expected, the first Bianchi identity results in the
homogeneous Maxwell equations. It is worth mentioning that in other
$5D$ theories, e.g. KK, equations (\ref{af}) or (\ref{ag}) do not
lead to equation (\ref{eq39}).

Now, let us see if our field equations, (\ref{ac1}), reduce to
Maxwell equations. If one, in the field equations replaces $A$ by
$4$ and $B$ by $\lambda$, one finds
\begin{equation} \hat{R}_{4\lambda}-\frac{1}{2}\hat{g}_{4\lambda}R=\kappa
\hat{T}_{4\lambda}. \label{eq50}
\end{equation}
 We can also see from equation
(\ref{eq33})  that
\begin{equation}
\hat{R}_{4 \lambda}=\hat{R}_{\lambda 4}=\hat{R}^{\alpha}_{\lambda \alpha
4}=\frac{1}{2}\nabla_{\alpha}F^{\alpha}_{\lambda}.
\end{equation}
Substituting this equation into (\ref{eq50}) yields the Maxwell
equations with a correction term
\begin{equation}
\nabla_{\alpha}F^{\alpha}_{\lambda}-\hat{g}_{4 \lambda}R=2\kappa
\hat{T}_{4\lambda},\label{ah}
\end{equation}
Note that in cases where the $4D$ scalar curvature is small, we may
neglect the correction term in (\ref{ah}). We can move $\hat{g}_{4
\lambda}R$ in (\ref{ah}) to the right hand side and look at it as a
new source. In the case of a large scalar curvature, this term,
under certain conditions, results in an electromagnetic field which
is stronger than that expected from the usual form of the Maxwell
equations. Recent observations may provide sufficient evidence
towards such a prediction. Indeed, it is possible to take account of
the unusually large galactic magnetic fields discovered recently
\cite{19,19a,19aa} for which no satisfactory explanation as yet
exists.
\section{Coordinate transformation}
If we change our coordinate system so that we again observe the space-time
from the $4D$ hypersurface, namely our universe, the main assumption
must be satisfied in order to yield the desired equations. It means
that we must have
\begin{equation}
\hat{\Gamma}^{4'}_{A'C'}=0.
\end{equation}
Therefore we need to examine the change of the above equation under
the transformation introduced as follows
\begin{equation}
\begin{array}{c}
~~~~~~~\frac{\partial \hat{x}^{4'}}{\partial\hat{x}^{4}}=const., \\
*\\
\frac{\partial \hat{x}^{4'}}{\partial\hat{x}^{\alpha}}=0\label{a9}.
\end{array}
\end{equation}
On the other hand we know that under transformations, the
connections change as
\begin{equation}
\hat{\Gamma}^{B'}_{A'E'}=\frac{\partial
\hat{x}^{B'}}{\partial\hat{x}^{C}}\left(\frac{\partial}{\partial\hat{x}^{A'}}\frac{\partial
\hat{x}^{C}}{\partial\hat{x}^{E'}}\right)+\frac{\partial
\hat{x}^{B'}}{\partial\hat{x}^{C}}\frac{\partial
\hat{x}^{D}}{\partial\hat{x}^{A'}}\frac{\partial
\hat{x}^{H}}{\partial\hat{x}^{E'}}\hat{\Gamma}^{C}_{DH}.
\end{equation}
If we now replace $B'$ by $4'$ in this equation
\begin{equation}
\hat{\Gamma}^{4'}_{A'E'}=\frac{\partial
\hat{x}^{4'}}{\partial\hat{x}^{C}}\left(\frac{\partial}{\partial\hat{x}^{A'}}\frac{\partial
\hat{x}^{C}}{\partial\hat{x}^{E'}}\right)+\frac{\partial
\hat{x}^{4'}}{\partial\hat{x}^{C}}\frac{\partial
\hat{x}^{D}}{\partial\hat{x}^{A'}}\frac{\partial
\hat{x}^{H}}{\partial\hat{x}^{E'}}\hat{\Gamma}^{C}_{DH},
\end{equation}
and substitute (\ref{a9}) into the above equation we find
\begin{equation}
\hat{\Gamma}^{4'}_{A'E'}=\frac{\partial \hat{x}^{D}}{\partial\hat{x}^{A'}}\frac{\partial \hat{x}^{H}}{\partial\hat{x}^{E'}}\hat{\Gamma}^{4}_{DH}.
\end{equation}
Hence, our assumption is invariant under the introduced
transformation.

\section{Conclusions} In this paper we have presented a unified
theory of gravity and electromagnetism where the resulting
inhomogeneous Maxwell equations are modified by a term involving the
curvature which, in certain cases, leads to electromagnetic fields
which are stronger than those obtained from the usual Maxwell
equations in cases where a large scalar curvature is present,
contrary to other conventional theories, e.g. Kaluza-Klein. This
could have interesting consequences and may be used to describe the
recently observed unusually strong galactic magnetic fields. This is
currently the focus of a work in progress.\vspace{3mm} \\
{\bf Acknowledgement}\vspace{1mm}\noindent\\
The authors would like to thank Professor Anzhong Wang for enlightening
discussions.

\end{document}